\documentclass[onecolumn]{emulateapj}
\newcommand       \be           {\begin{equation}}
\newcommand       \ee           {\end{equation}}

\newcommand       \bomega       {\mbox{\boldmath$\omega$\unboldmath}}

\shortauthors{Weingartner}
\shorttitle{Thermal Flipping of Interstellar Grains}

\begin{document}

\title{Thermal Flipping of Interstellar Grains}

\author{Joseph C. Weingartner}
\affil{Department of Physics and Astronomy, George Mason University,
MSN 3F3, 4400 University Drive, Fairfax, VA 22030, USA}
\email{joe@physics.gmu.edu}

\begin{abstract}

In interstellar dust grains, internal processes dissipate rotational kinetic 
energy.  The dissipation is accompanied by thermal fluctuations,
which transfer energy from the vibrational modes to rotation.  Together,
these processes are known as internal relaxation.  For the past several
years, internal relaxation has been thought to give rise to thermal 
flipping, with profound consequences for grain alignment theory.  I show
that thermal flipping is not possible 
in the limit that the inertia tensor does not vary with time.

\end{abstract}


\section{\label{sec:intro} Introduction}

Two processes act in concert to align grains with the interstellar magnetic
field:  (1)  the grain's principal axis of greatest moment of inertia 
$\hat{a}_1$ aligns with respect to its angular momentum vector $\mathbf{J}$
and (2)  $\mathbf{J}$ aligns with respect to the magnetic field vector
$\mathbf{B}$.  Purcell (1979) noted that internal mechanisms for dissipating
rotational energy drive the grain to its lowest energy state for a given 
$\mathbf{J}$, namely steady rotation with $\hat{a}_1 \parallel \mathbf{J}$.  
This occurs on a much shorter timescale than that on which external 
processes align $\mathbf{J}$ relative to $\mathbf{B}$.  

Purcell (1979) identified two internal dissipation mechanisms.  Inelastic
dissipation results from the periodic mechanical 
stresses experienced by a grain that does not 
rotate steadily about a principal axis.  The existence of this process 
is fairly obvious, but calculating the dissipation rate is  
a challenging problem (see, e.g., Sharma et al. 2005 and references therein).  
Purcell (1979) introduced
a second, subtle effect, which he termed ``Barnett dissipation''.  
When a grain does not rotate steadily about a principal axis, the angular
velocity vector $\bomega$
varies periodically in a coordinate system attached to the
grain.  If the grain consists of a paramagnetic material, then the 
microscopic spins (with gyromagnetic ratio $\gamma_g$) 
attempt to align with the fictitious
``Barnett-equivalent'' 
magnetic field $\mathbf{B}_{\rm BE} = \bomega / \gamma_g$.  As
the grain magnetization attempts to follow $\mathbf{B}_{\rm BE}$, rotational
kinetic energy is dissipated.  This process is analogous to a magnetic
resonance experiment, where the dissipated energy is provided instead by the 
applied radiation field.

Purcell (1979) provided a heuristic derivation of the Barnett dissipation
rate for oblate 
grains with dynamic symmetry.  ``Dynamic symmetry'' refers to the case that 
$I_2 = I_3$, 
where $I_i$ are the moments of inertia associated with the principal axes
$\hat{a}_i$.  Thus, for oblate grains with this symmetry, 
$I_1 > I_2 = I_3$.  
(Henceforth, the term ``oblate'' shall always refer to dynamic,
rather than geometric, symmetry.)  
In this case, $\mathbf{B}_{\rm BE}$ consists of a static 
component $(\omega_{\parallel}/\gamma_g) \hat{a}_1$ plus a component 
$\mathbf{B}_{\rm BE, \, rot}$ with 
magnitude $\omega_{\perp}/\gamma_g$ that rotates in the 
$\hat{a}_2-\hat{a}_3$ plane with angular speed $\omega_{\rm rot}$. 
Solving the Euler equations yields
\be
\label{eq:omega_parallel_oblate}
\omega_{\parallel} = \frac{J}{I_1} \cos \gamma~~~,
\ee
\be
\omega_{\perp} = \frac{J}{I_2} \sin \gamma~~~,
\ee
and
\be
\omega_{\rm rot} = \frac{J(I_1-I_2)}{I_1 I_2} \cos \gamma~~~,
\ee
where $\gamma$ is the (constant) angle between $\mathbf{J}$ and $\hat{a}_1$.
Assuming $\mathbf{J}$ and $I_1$ are constant, 
it is 
convenient to introduce a dimensionless measure of the rotational energy
$E$:
\be
\label{eq:q}
q \equiv \frac{2 I_1 E}{J^2} = 1 + (r_2 - 1) \sin^2 \gamma~~~,
\ee
where the final equality is for oblate grains.  
Note that, for oblate grains, $q$ ranges from 1 to $r_2 \equiv I_1 / I_2$. 

Purcell (1979) argued that the dissipation rate is given by 
\be
\label{eq:Barnett_diss}
\left( \frac{dE}{dt} \right)_{\rm Bar} = - V \chi^{\prime \prime}
B_{\rm BE, \, rot}^2 \omega_{\rm rot}~~~,
\ee
where $V$ is the grain volume and $\chi^{\prime \prime}$ is the
imaginary component of the magnetic susceptibility.
It is worth noting that, although this expression (as well as a variant in 
Lazarian \& Draine 1999b) is widely used in grain alignment theory, it has
not yet been rigorously derived or experimentally verified.  
Purcell adopted the low-frequency susceptibility
\be
\label{eq:chi_imag}
\chi^{\prime \prime} \approx \chi_0 \omega_{\rm rot} T_2~~~,
\ee
where $\chi_0$ is the static susceptibility and $T_2$ is the
spin-spin relaxation time.  

Combining equations (\ref{eq:omega_parallel_oblate}) through 
(\ref{eq:chi_imag}) yields
\be
\label{eq:dqdt_low_J}
\frac{dq}{dt} = - \tau_{\rm Bar}^{-1} (q-1) (r_2 - q)
\ee
with
\be
\label{eq:tau_Bar}
\tau_{\rm Bar} = \frac{\gamma_g^2 I_1 I_2^2}{2 \chi_0 V T_2 J^2}~~~.
\ee
More realistic approximations for $\chi^{\prime \prime}$ (e.g., Draine \& 
Lazarian 1999) yield more complicated expressions for $dq/dt$, but retain
the linear dependence on $(q-1)$ and $(r_2-q)$ near $q=1$ and $q=r_2$, 
respectively.  

In the inverse process of Barnett dissipation, a fluctuation spontaneously 
transfers some energy from the thermal reservoir provided by the grain 
vibrational modes to the grain rotation.  Lazarian \& Draine (1997) showed
that these thermal fluctuations can play an important role in grain alignment. 
They examined the classic alignment model, developed by Purcell (1975, 1979), 
in which a systematic torque $\mathbf{\Gamma}_{\rm sys}$, fixed in grain body 
coordinates, spins the grain to ``suprathermal'' rotation.  Thermal rotation,  
arising solely from collisions with particles from a gas with temperature 
$T_{\rm gas}$, is characterized by 
$J \sim J_{\rm th} \equiv \sqrt{I_1 k_B T_{\rm gas}}$
($k_B$ is Boltzmann's constant).
Suprathermally-rotating grains, with $J \gg J_{\rm th}$, are impervious to 
disalignment by random collisions with gas atoms.
Thus, $\mathbf{J}$ can gradually align with $\mathbf{B}$ via the 
Davis-Greenstein (1951) mechanism.  Purcell (1979) found that the most 
important systematic torque results from the formation (and subsequent 
ejection) of H$_2$ molecules at special sites on the grain surface.  

The distribution of molecule-forming surface sites can change rapidly 
compared with the Davis-Greenstein alignment rate (see, e.g., Lazarian 
1995).    
As a result of this resurfacing, $\mathbf{\Gamma}_{\rm sys} \cdot \hat{a}_1$
may reverse sign, sometimes spinning the grain down to 
thermal rotation.  (In inertial coordinates, $\mathbf{\Gamma}_{\rm sys}$
is always parallel or 
anti-parallel to $\mathbf{J}$, when averaged over the grain rotation.)
These episodes, known as crossovers, were first studied
by Spitzer \& McGlynn (1979), who concluded that thorough disalignment of
$\mathbf{J}$ relative to $\mathbf{B}$ 
occurs after passage through a small number of crossovers.  Lazarian \& 
Draine (1997) found that the small disalignment of $\hat{a}_1$ from 
$\mathbf{J}$ during periods of suprathermal rotation (due to thermal
Barnett 
fluctuations) limits the minimum value of $J$ during a crossover, thereby 
limiting the disalignment of $\mathbf{J}$ with $\mathbf{B}$.  

Although Lazarian \& Draine (1997) found that thermal fluctuations during
periods of suprathermal rotation may aid grain alignment, these same authors
soon concluded that thermal fluctuations during periods of slow rotation
may severely suppress alignment.  
Lazarian \& Draine (1999a) introduced the concepts of thermal flipping and
thermal trapping.  
For an oblate grain, internal thermal fluctuations cause the angle $\gamma$ 
between $\mathbf{J}$ and $\hat{a}_1$ to vary stochastically (see 
eq.~\ref{eq:q}).  Whenever $\gamma$ crosses
$\pi/2$ (a ``thermal flip''), 
$\mathbf{\Gamma}_{\rm sys} \cdot \mathbf{J}$ changes
sign.  If flips occur rapidly, then the grain can only achieve suprathermal
rotation if it reaches, by random walk, a sufficiently high $J$ that the 
flipping timescale (which increases with $J$) exceeds the spin-up timescale.
Grains for which suprathermal rotation is thereby suppressed are 
``thermally trapped''.  

Purcell (1979) considered only the contribution of electron paramagnetism
to Barnett dissipation.  Lazarian \& Draine (1999b) showed that nuclear
paramagnetism can yield much larger dissipation rates for thermally-rotating
grains.  When including ``nuclear relaxation'' in their analysis, they found 
that grains with size up to $1 \micron$ are trapped.
Thus, the Purcell (1979) scenario of Davis-Greenstein alignment of 
suprathermally rotating grains appears to fail, unless the grains contain
superparamagnetic inclusions (Jones \& Spitzer 1967).  

Radiative torques (Harwit 1970a, b; Dolginov 1972; Draine \& Weingartner 
1996, 1997; Weingartner \& Draine 2003; Hoang \& Lazarian 2008;
Lazarian \& Hoang 2007, 2008), 
which are not fixed in grain body coordinates, have the potential to rapidly 
align $\mathbf{J}$ with $\mathbf{B}$.  Weingartner \& Draine (2003) 
found that radiative torques can drive grains into various alignment states,
some characterized by thermal rotation and some by suprathermal rotation.
For the former states, the grains were thought to undergo rapid flipping.
This result was confirmed by additional calculations in Hoang \&
Lazarian (2008), who also noted that thermally rotating, aligned grains
may ultimately reach aligned states characterized by
suprathermal rotation, due to random gas
atom impacts.  

Thermal flipping appears to play a critical role in grain alignment theory,
precluding the Purcell scenario (i.e., Davis-Greenstein alignment with 
suprathermal rotation suppressing disalignment) 
and affecting the aligned grain states in 
the radiative torque scenario.  Thus, a quantitative estimate of the flipping
rate is needed.  This can be accomplished with the use of the Langevin 
and/or Fokker-Planck equations.  (Gardiner 2004 provides an excellent 
introduction to stochastic methods.)  In \S 2, I will 
show that thermal flipping as described by Lazarian \& Draine (1999a) is,
in fact, not possible.  

\section{\label{sec:langevin_eqn} The Langevin Equation for Internal 
Relaxation}

The Langevin equation is a stochastic differential equation describing
the time evolution of the grain rotational energy:
\be
\label{eq:langevin}
dq = A(q) \, dt + \sqrt{D(q)} \, dw~~~,
\ee
where $dw$ is a Gaussian random variable with variance $dt$.  
For Barnett dissipation (in the Purcell 1979 approximation), 
the drift coefficient $A(q)$ is given by the right hand side of equation
(\ref{eq:dqdt_low_J}).  Ideally, the diffusion 
coefficient $D(q)$ would also be derived from the model for Barnett 
relaxation, but no model has been developed with sufficient detail for
this to be possible.  Instead, $D(q)$ can be determined (to within a 
constant of integration) by demanding that
the probability current $S(q)$ vanish at all $q$ when thermal 
equilibrium obtains.  If $f(q) dq$ is the probability that the dimensionless
energy lies between $q$ and $q+dq$, then 
\be
\label{eq:prob_current}
S(q) = A f - \frac{1}{2} \frac{d(fD)}{dq}
\ee
(eq.~5.2.8 in Gardiner 2004).  

Weingartner \& Draine (2003) defined the quantity
\be
\label{eq:s}
s \equiv 1 - \frac{2}{\pi} \int_0^{\alpha_{\rm max}} d\alpha \left[
\frac{I_3 (I_1 - I_2 q) + I_1 (I_2 - I_3) \cos^2
\alpha}{I_3 (I_1 - I_2) + I_1 (I_2 - I_3) \cos^2 \alpha} \right]^{1/2}~~~,
\ee
where
\be
\alpha_{\rm max} \equiv \cases{\pi/2 &, $q \le I_1/I_2$\cr
\cos^{-1} \left[ \frac{I_3 (I_2 q -I_1)}{I_1 (I_2 - I_3)} \right]^{1/2}
&, $q > I_1/I_2$\cr}~~~,
\ee
and showed that the density of energy states is constant in $s$.  This
holds for grains with arbitrary $I_1$, $I_2$, $I_3$.  For oblate grains
($I_2 = I_3$),
\be
s = 1 - \left( \frac{r_2 - q}{r_2 - 1} \right)^{1/2}~~~.
\ee
Thus, for oblate grains, the thermal equilibrium distribution function is
\be
f_{\rm TE}(q) \propto \exp(-kq) \frac{ds}{dq} \propto \exp(-kq)
(r_2 - q)^{-1/2}~~~,
\ee
where
\be
k \equiv \frac{J^2}{2 I_1 k_B T_d}~~~.
\ee
The thermal equilibrium distribution function is more complicated for grains 
lacking dynamic symmetry, but still depends on $k$.  

The impossibility of thermal flipping can be simply demonstrated 
by examining the relaxation at $q=r_2$ in the limit that the dust temperature
$T_d \rightarrow 0$.  In this limit, 
$k \rightarrow \infty$.  As $T_d \rightarrow 0$, fluctuations cease to
contribute to the probability current $S(q)$, implying that
$d(fD)/dq \rightarrow 0$.  This limiting behavior must hold for all $q$
(including $q=r_2$) and for any physically realizable probability
distribution $f(q)$.  Normalization of $f(q)$ requires that any divergence
at $q=r_2$ be shallower than $f(q) \propto (r_2 -q)^{-1}$, unless
$f(q) = \delta(q-r_2)$.  

Of course, the contribution of drift to the current,
$A(q) f(q)$, must also vanish at $q=r_2$.  Evidently, it is necessary that
$A(q)$ falls off linearly or faster with $(r_2-q)$ near $q=r_2$.  Note that
the Barnett dissipation rate of equation
(\ref{eq:dqdt_low_J}) does satisfy this condition.  
Equation (\ref{eq:tau_Bar}) suggests that $\tau_{\rm Bar} \rightarrow 0$ as 
$T_d \rightarrow 0$, since $\chi_0 \propto T_d^{-1}$.  However, this cannot
be correct, since $A(q = r_2)$ would be undefined rather than
zero as $T_d \rightarrow 0$.   
Equation (\ref{eq:tau_Bar}) does not hold for $T_d$ lower than the
Curie temperature.  For such low temperatures, the material is ferromagnetic,
suggesting that $\tau_{\rm Bar}$ approaches a non-zero constant as 
$T_d \rightarrow 0$.  

Focusing now on the fluctuating term at $q=r_2$,
\be
\label{eq:lim}
\lim_{(k^{-1}, r_2-q) \rightarrow (0,0)} \frac{d[f(k, q) D(k,q)]}{dq} = 0~~~.
\ee
The limit only exists if it takes the same value for all paths along which 
$(k^{-1}, r_2-q) \rightarrow (0,0)$.  Since there exist paths for which 
$(r_2-q) \rightarrow 0$ arbitrarily more rapidly than $k^{-1} \rightarrow 0$, 
$d(fD)/dq$ may not contain any divergences with respect to $q=r_2$.  
Thus, for $(r_2 -q) \ll 1$, $fD$ must either (1) be independent of $(r_2-q)$ 
or (2) fall off linearly or faster with $(r_2-q)$.  If condition (1)
holds for a particular distribution $f_1(q)$, then it will not hold for
another distribution $f_2(q)$ having a different dependence on $(r_2-q)$ near 
$q=r_2$.  Thus, condition (2) must generally obtain, implying that
$D$ must fall off as $(r_2-q)^2$ or faster near
$q=r_2$.  This implies that $D$ and $dD/dq$ both vanish at $q=r_2$.
If the diffusion coefficient is smooth with respect to $k$
(in the sense that $dD/dk$ exists for all $k$), then these conditions must
be satisfied for all $k$.  

Thus, $A$, $D$, and $dD/dq$ all vanish at $q=r_2$, making this point a
``natural boundary'' (see \S 5.2.1e of Gardiner 2004).
A system can never reach a natural boundary if it begins at a different
point (i.e., with a different value of $q$).  However, the system must
reach $q=r_2$ and return to lower $q$ (with a different sign
for $\cos \gamma$) in order for a flip to occur (see \S 2.5.2 of Weingartner 
\& Draine 2003).  Consequently, thermal 
flipping is prohibited.  This conclusion does not
depend on the form of $A(q)$, except that $A(q)$ decreases as $(r_2-q)$ or
faster for $q$ near $r_2$.  It holds for any type of internal relaxation
and for grains with or without dynamic symmetry, so long as $dD/dk$ exists
for all $k$.  

If thermal flipping is truly prohibited, then this result must obtain 
regardless of the choice of variable.  Although the current $S$ is 
independent of variable, the two terms composing it, representing drift
and diffusion, are not.  
When transforming variables in stochastic differential equations, the
ordinary rules of calculus only apply for linear transformations.  Otherwise,
Ito's formula must be used (see \S 4.3.3 of Gardiner 2004).  When the
Langevin equation (\ref{eq:langevin}) is transformed to variable $y(q)$,
the result is
\be
\label{eq:ito}
dy = \left[ A(q) \frac{dy}{dq} + \frac{1}{2} D(q) \frac{d^2 y}{dq^2} \right] dt
+ \sqrt{D(q)} \frac{dy}{dq} dw~~~.
\ee
Note the additional contribution to the drift coefficient when the Langevin
equation is written in the new variable.
Ito's formula, along with the relation $f(q) dq = f(y) dy$, yields
\be
\label{eq:diff_term}
\frac{d}{dy} \left[ f(y) D(y) \right] = \frac{d}{dq} \left[ f(q) D(q)
\right] + f(q) D(q) \frac{d^2 y/dq^2}{dy/dq}~~~.
\ee
If $y(q) \propto (r_2 -q)^p$ (with $p \ne 0$) for $q$ near $r_2$, then 
\be
\frac{d^2 y/dq^2}{dy/dq} \propto (r_2 -q)^{-1}~~~.
\ee
Since $D(q) f(q) \propto (r_2 -q)^n$ with $n > 1$, the second
term in equation (\ref{eq:diff_term}) vanishes at $q=r_2$.  Thus, if the 
diffusion 
contribution to the current vanishes at the point $q=r_2$ for variable $q$, 
then it does so for arbitrary variable.  

To illustrate the above arguments in a concrete setting, I will now 
discuss the diffusion coefficient $D(q)$ for an oblate grain and the 
approximate dissipation rate of equation (\ref{eq:dqdt_low_J}).  
Setting the probability current equal to zero for thermal equilibrium yields
\be
\label{eq:D_q}
D(q) = \frac{1}{f_{\rm TE}(q)} \left[ D(1) f_{\rm TE}(1) + 2
\int_1^q A(q^{\prime}) f_{\rm TE}(q^{\prime}) dq^{\prime} \right]~~~.
\ee
Upon integrating, 
\begin{eqnarray}
\nonumber
\label{eq:D}
k^2 \tau_{\rm Bar} D(q) & = & [3 + 2 k (q-1)] (r_2 - q) + C(k) (r_2 - q)^{1/2}
\exp(kq)\\
& &  - k^{-1/2} [3 + 2 k (r_2 -1)] (r_2 - q)^{1/2} \exp[-k(r_2 -q)]
\int_0^{\sqrt{k (r_2 -q)}} \exp(x^2) dx
\end{eqnarray}
with
\begin{eqnarray}
\nonumber
C(k) & = & k^2 \exp(-k) (r_2 -1)^{-1/2} D(q=1, k) \tau_{\rm Bar} - 3 \exp(-k)
(r_2 -1)^{1/2}\\
& &  + k^{-1/2} [3 + 2 k (r_2 -1)] \exp(-k r_2)
\int_0^{\sqrt{k (r_2 -1)}} \exp(x^2) dx~~~.
\label{eq:C}
\end{eqnarray}

For $(r_2 - q) \ll 1$,
\be
k^2 \tau_{\rm Bar} D(q, k) \approx C(k) \exp(kq) (r_2 - q)^{1/2} +
\frac{4}{3} k^2 (r_2 -1) (r_2 -q)^2~~~~,~~(r_2 - q) \ll 1~~~.
\ee
The term containing $C(k)$ does not fall off sufficiently quickly with
$(r_2 -q)$.  Thus, $C(k) = 0$ identically (for all $k$, if $dD/dk$ exists
for all $k$).  The remaining term varies as $(r_2-q)^2$, the shallowest
permissible dependence.  Note that the term containing $C(k)$ satisfies
condition (1) following equation (\ref{eq:lim}) for the thermal equilibrium
distribution $f_{\rm TE}(q)$, but not the required general condition (2).

Given the above general argument prohibiting thermal flipping induced by
internal relaxation, one may ask how Lazarian \& Draine (1999a) concluded
that thermal flipping is possible.  Their analysis was highly approximate
and did not employ any diffusion coefficient.  Nevertheless, their 
estimate of the flipping rate agreed well with the detailed analysis of
Roberge \& Ford (1999), which made use of the Barnett relaxation
diffusion coefficient calculated by Lazarian \& Roberge (1997).  

Lazarian \& Roberge (1997) solved a modified version of equation 
(\ref{eq:prob_current}) for the diffusion coefficient, in which they 
used the angle $\gamma$ rather than $q$ as the variable.  They considered
oblate grains and the approximate Barnett dissipation rate
in equation (\ref{eq:dqdt_low_J}).  Using equation (\ref{eq:q}) to
substitute for $q$ in terms of $\gamma$ in equation (\ref{eq:dqdt_low_J}), 
they adopted
\be
\label{eq:A_LR97}
A(\gamma) = - \frac{r_2 -1}{2 \tau_{\rm Bar}} \sin \gamma \cos \gamma
\ee
(see their eqs.~1, 2, 4, and 16).  
In Purcell's (1979) heuristic derivation of the Barnett dissipation rate,
he obtained the rate at which the rotational energy $E$ decreases. 
In other words, he obtained the drift coefficient $A(E)$.  
Since $q \propto E$, there is no additional contribution to the drift 
coefficient arising from Ito's formula (\ref{eq:ito}) when the 
Langevin equation is written in the variable $q$.  However, the variable
in Lazarian \& Roberge (1997) is the angle $\gamma$, which is a non-linear
function of $E$ (eq.~\ref{eq:q}).  The additional contribution to
the drift coefficient was not included in their analysis.  Despite this
error, the above general argument should still yield vanishing $D$ and 
$dD/d\gamma$ at $\gamma = \pi/2$, and thus no thermal flipping.  

In the vicinity of $\gamma = \pi/2$ (corresponding to $q=1$), the 
diffusion coefficient calculated by Lazarian \& Roberge (1997; their 
eq.~18) is
\be
\tau_{\rm Bar} D(\gamma) = \tau_{\rm Bar} D(\gamma = \pi/2) + \left\{ 1 - 
\left[ \left(r_2 -1 \right) k - \frac{1}{2} \right] \tau_{\rm Bar}
D(\gamma = \pi/2) \right\} (\gamma -\pi/2)^2~~~.
\ee
If $D(\gamma = \pi/2)$ is taken to be zero, then 
$D \propto (\gamma -\pi/2)^2$, as required.  In this case, thermal flipping
does not occur.  Lazarian \& Roberge (1997) argued that $D$ should be 
smooth with respect to $\gamma$ and demanded that $d^2 D/d\gamma^2$ exist for 
all $k$.  This condition is satisfied if $D(\gamma = \pi/2) = 0$ or if
$D \propto k^n$ with $n \le -1$.  Lazarian \& Roberge (1997) chose 
$D \propto k^{-1}$, which admits thermal flipping but is inconsistent with the
requirement that $D(\gamma)$ falls off at least as quickly as 
$(\gamma -\pi/2)^2$ for $\gamma$ near $\pi/2$.  [There appear to be some
typographical errors in Lazarian \& Roberge 1997.  In their eq.~19, 
$D \propto k^{-1/2}$ rather than $k^{-1}$.  In their eq.~18, 
$D(\gamma = \pi/2) = 1$ rather than falling off as $k^{-1/2}$ or $k^{-1}$.]

Lazarian \& Roberge (1997) tested their result for the diffusion coefficient
by numerically evolving their Langevin equation for a large number of 
Barnett timescales and computing the average value of the internal 
alignment factor
\be
\label{eq:Q_X}
Q_X \equiv \frac{3}{2} \left[ \langle \cos^2 \gamma \rangle - \frac{1}{3}
\right]~~~.
\ee
This can also be evaluated by simple integration for a thermal distribution
(their eq.~10).  The results of their simulations agreed to high accuracy 
with the direct calculations.  They adopted the wrong Langevin equation
but the correct thermal equilibrium 
distribution function.  Their success with the test
indicated that they solved equation (\ref{eq:prob_current}) correctly
given their drift coefficient, but this drift coefficient does not
describe Barnett dissipation when angle $\gamma$ is taken as the variable.  

As a confidence-building check on the conclusion that thermal flipping is 
prohibited, I numerically evolved
the Langevin equation for the case that $k=1$ and $r_2 = 1.5$, taking 
$C=0$.  A fixed time step size is attempted at each step.  Sometimes
this results in overshooting $q=1$; in these cases, smaller steps are
tried until the resulting $q$ exceeds 1.  These overshooting incidents
become fractionally less common as the base step size is decreased (from 
$10^{-2} \tau_{\rm Bar}$ to $10^{-5} \tau_{\rm Bar}$).  The total duration
of a simulation is about $10^5 \tau_{\rm Bar}$.  At no time, for any 
of the base step sizes, did $q$ ever overshoot $r_2$.  Incidentally, the
simulations yielded the correct value for the alignment factor $Q_X$, although 
the convergence was slower
than for the simulations in Lazarian \& Roberge (1997).

The above argument that $D$ and $dD/dq$ both vanish at $q=r_2$ made no 
reference to the form of $A(q)$.  Thus, this conclusion also holds when  
more realistic Barnett dissipation rates are adopted,
and even for grains lacking dynamic symmetry. In all of these cases, 
$A(q=r_2) = 0$, since the grain is in steady rotation when $q=r_2$.  
Thus, $q=r_2$ is a natural boundary for the most general treatment of 
Barnett relaxation, if $dD/dk$ exists for all $k$.  Since a grain lacking
dynamic symmetry must reach (and, in general, cross) 
$q=r_2$ in order to flip (see \S 2.5.2 of Weingartner \& Draine 2003), 
thermal flipping associated with Barnett relaxation is ruled out 
generally.  The only caveat is that $dD/dk$ must exist for all $k$.  
Although this seems natural, a detailed model of Barnett relaxation would
be needed to confirm that this condition is indeed satisfied.  

The discussion here has focused on Barnett relaxation, since Barnett
dissipation appears to dominate inelastic dissipation for most thermally
rotating grains (Lazarian \& Efroimsky 1999), especially when nuclear
paramagnetism contributes.  However, the argument against thermal flipping
applies equally well for inelastic relaxation.  Lazarian \& 
Efroimsky (1999) did not constrain the form of the dissipation rate 
near $q=r_2$, but Sharma et al.~(2005) found the same form as in 
equation (\ref{eq:dqdt_low_J}) for the special case of an oblate spheroid.

\section{Conclusion}

In conclusion, it appears that thermal flipping is not possible, so long as
$dD/dk$ exists for all $k$ and the inertia tensor does not vary with time.    
A detailed model of Barnett relaxation is needed to examine the behavior
of $dD/dk$.  

Because of grain vibrations, the inertia tensor exhibits continual, small 
variations.  As a result, the location of the natural boundary at 
$q = r_2 \equiv I_1/I_2$ wanders slightly (B.T. Draine, private communication).
Further work is needed to examine whether this can give rise to flips and, if
so, at what rate.  

External processes (e.g., gas atom impacts) may
also induce flips (with accompanying changes in $\mathbf{J}$), but have 
recently been neglected in comparision with
internal relaxation (e.g., Weingartner \& Draine 2003).  These now merit
further scrutiny as well.

\acknowledgements

I am grateful to Bruce Draine, Wayne Roberge, and Alex Lazarian for 
illuminating discussions and comments on the manuscript.  
JCW is a Cottrell Scholar of Research Corporation.
Support for this work, part of the Spitzer Space Telescope Theoretical
Research Program, was provided by NASA through a contract issued by the
Jet Propulsion Laboratory, California Institute of Technology under a
contract with NASA.


\begin{thebibliography}

\bibitem[]{DG51} Davis, L. \& Greenstein, J. L. 1951, \apj, 114, 206
\bibitem[]{Do72} Dolginov, A.Z. 1972, Ap\&SS, 18, 337
\bibitem[]{DL99} Draine, B. T. \& Lazarian, A. 1999, \apj, 512, 740
\bibitem[]{DW96} Draine, B. T. \& Weingartner, J. C. 1996, \apj, 470, 551
\bibitem[]{DW97} Draine, B. T. \& Weingartner, J. C. 1997, \apj, 480, 633
\bibitem[]{G04} Gardiner, C. W. 2004, Handbook of Stochastic Methods, 3ed
(Berlin: Springer)
\bibitem[]{Har70a} Harwit, M. 1970a, Nature, 226, 61
\bibitem[]{Har70b} Harwit, M. 1970b, Bull. Astron. Inst. Czechoslovakia,
21, 204
\bibitem[]{HL08} Hoang, T. \& Lazarian, A. 2008, \mnras, 388, 117
\bibitem[]{JS67} Jones, R. V. \& Spitzer, L. 1967, \apj, 147, 943
\bibitem[]{L95} Lazarian, A. 1995, \mnras, 274, 679
\bibitem[]{LD98} Lazarian, A. \& Draine, B. T. 1997, \apj, 487, 248
\bibitem[]{LD99a} Lazarian, A. \& Draine, B. T. 1999a, \apj, 516, L37
\bibitem[]{LD99b} Lazarian, A. \& Draine, B. T. 1999b, \apj, 520, L67
\bibitem[]{LE99} Lazarian, A. \& Efroimksy, M. 1999, \mnras, 303, 673
\bibitem[]{LH07} Lazarian, A. \& Hoang, T. 2007, \mnras, 378, 910
\bibitem[]{LH08} Lazarian, A. \& Hoang, T. 2008, \apj, 676, L25
\bibitem[]{LR97} Lazarian, A. \& Roberge, W. G. 1997, \apj, 484, 230
\bibitem[]{P75} Purcell, E. M. 1975, in The Dusty Universe, ed. G. B. Field
\& A. G. W. Cameron (New York:  Neal Watson), 155
\bibitem[]{P79} Purcell, E. M. 1979, \apj, 231, 404
\bibitem[]{RF99} Roberge, W.G. \& Ford, K.E.S. 1999, unpublished 
\bibitem[]{SBH05} Sharma, I., Burns, J. A., \& Hui, C.-Y. 2005, \mnras, 359, 79
\bibitem[]{SM79} Spitzer, L. \& McGlynn, T. A. 1979, \apj, 231, 417
\bibitem[]{WD03} Weingartner, J. C. \& Draine, B. T. 2003, \apj, 589, 289






\end{thebibliography}
\end{document}